\documentclass[%
 reprint,
 amsmath,amssymb,
 aps,
]{revtex4-1}
\usepackage{siunitx}
\usepackage{graphicx}
\usepackage{dcolumn}
\usepackage{bm}
\usepackage{color}

\begin{document}

\preprint{APS/123-QED}

\title{A two-way photonic quantum entanglement transfer interface}

\author{ Yiwen Huang$^{1,\ddagger}$, Yuanhua Li$^{1,2,\ddagger}$}%
 \email{lyhua1984@sjtu.edu.cn}
\author{Zhantong Qi$^{1}$, Juan Feng$^{1}$, Yuanlin Zheng$^{1,3}$}

\author{Xianfeng Chen$^{1,3,4,5}$}%
 \email{xfchen@sjtu.edu.cn}
\affiliation{%
 $^1$ State Key Laboratory of Advanced Optical Communication Systems and Networks, School of Physics and Astronomy, Shanghai Jiao Tong University, Shanghai 200240, China\\ $^2$ Department of Physics, Jiangxi Normal University, Nanchang 330022, China\\ $^3$ Shanghai Research Center for Quantum Sciences, Shanghai 201315, China\\ $^4$ Jinan Institute of Quantum Technology, Jinan 250101, China\\ $^5$ Collaborative Innovation Center of Light Manipulation and Applications, Shandong Normal University, Jinan 250358, China\\$\ddagger$ These authors contributed equally to this work
}%


\date{\today}

\begin{abstract}
 A quantum interface for two-way entanglement transfer between orbital angular momentum degree of freedom in free space and time-energy degree of freedom in optical fibers, provides a novel way toward establishing entanglement between remote heterogeneous quantum nodes. Here, we experimentally demonstrate this kind of transfer interface by using two interferometric cyclic gates. By using this quantum interface, we perform two-way entanglement transfer for the two degrees of freedom. The results show that the quantum entangled state can be switched back and forth between orbital angular momentum and time-energy degrees of freedom, and the fidelity of the state before and after switching is higher than 90\%. Our work demonstrates the feasibility and high performance of our proposed transfer interface, and paves a route toward building a large-scale quantum communication network.

\end{abstract}

\maketitle


\textit{Introduction}-An entanglement-based quantum network is a platform for the science and application of secure communication and distributed quantum computation. In a complex network, quantum entanglement can be encoded in various degrees of freedom (DOF), such as polarization, time-energy and orbital angular momentum (OAM). Due to the unique phase-intensity profile and unlimited number of orthogonal modes, OAM entangled states have engendered a variety of quantum applications \cite{erhard2018twisted}, such as high-dimensional quantum key distribution \cite{mafu2013higher}, quantum teleportation \cite{wang2015quantum}, high-dimensional entanglement swapping\cite{Zhang2017Simultaneous}, fundamental tests of quantum mechanics \cite{dada2011experimental}, digital spiral imaging \cite{chen2014quantum}, quantum pattern recognition \cite{qiu2019structured} and transmission matrix measurement \cite{valencia2020unscrambling}. OAM entanglement is currently more and more widely used in quantum communication tasks. On the other hand, time-energy entanglement is of great interest as it supports various encodings \cite{MacleanDirect,martin2017quantifying} and is insensitive to the birefringence effect of fibers \cite{Marcikic2002,Zhang2008,Grassani2015}. Different from polarization entanglement which requires real-time active control to compensate polarization drifts \cite{Treiber2009,Yu2015,Wengerowsky2018,Joshieaba0959}, time-energy entanglement, both continuous and discrete versions, shows intrinsic robustness for propagation through long-distance fiber with the help of passive dispersion-compensating devices \cite{Lee2014,Aktas2016}.  To date, time-energy entanglement sources have been widely used in optical-fiber quantum tasks, such as quantum key distribution \cite{zhong2015photon,Zhang2014unconditional,Liu2020}, dense coding \cite{Williams2017Superdense} and long-distance quantum entanglement swapping \cite{li2019multiuser}. It is an important candidate for building long-distance optical fiber networks.

The future quantum communication network is composed of free space and optical-fiber coupling connection. In order to accomplish different quantum tasks, the nodes of the network need to transfer information-carrying entangled photons back and forth in free space and optical fibers. This requires a reversible quantum entanglement transfer (QET) interface to perfectly switch the entangled photons back and forth in free space and optical fibers, which is the core technology for realizing quantum communication between nodes in the network. So far, QET has been implemented from time to polarization \cite{Vasconcelos2020}, polarization to OAM \cite{Nagali2009}, and polarization to frequency \cite{Ramelow2009} DOF. Implement of a two-way QET interface that can control quantum entanglement switching back and forth between time-energy and OAM DOF is urgently needed for constructing the large-scale quantum communication network. However, such a two-way QET interface of entangled photons has not been implemented.

Here we demonstrate the first experiment of QET between time-energy and OAM DOF of photons. Two interferometric quantum gates consisting of a Franson-type interferemeter with spiral phase plates (SPP) inserted in different paths are utilized for transferring quantum entanglement information from time-energy to OAM DOF. Furthermore, we use two OAM sorters followed by two Mach-Zehnder interferometers (MZIs) to implement QET from OAM to time-energy DOF. The experiment results reveal a high quality of the QET between these two DOF, while preserving quantum coherence and entanglement. Our approach paves the way towards a novel means of connecting remote heterogeneous quantum nodes in time-energy and OAM subspace.

\begin{figure*}[htbp]
\centering
\includegraphics[width=0.9\linewidth]{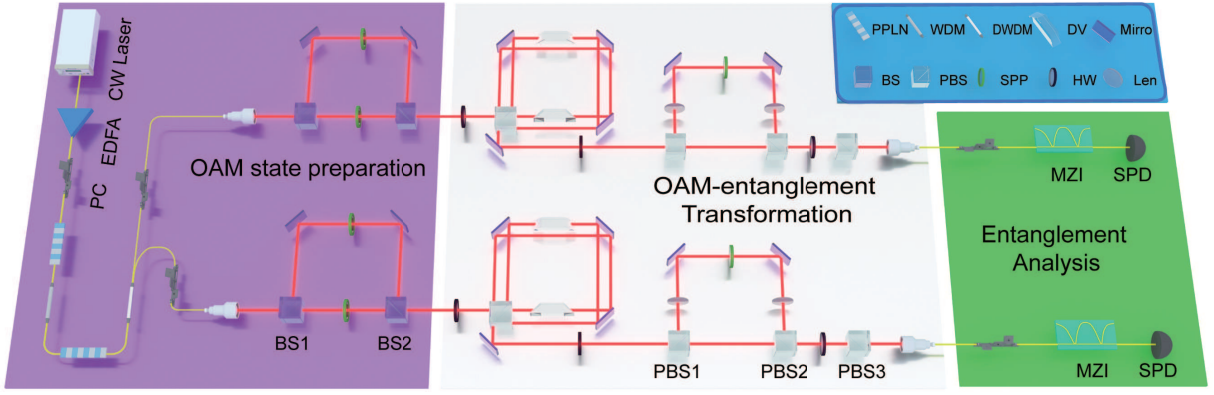}
\caption{Experimental setup of QET. All the MZIs in OAM state preparation and transformation part are steadily phase-locked by PZT systems. EDFA, erbium doped fiber amplifier; WDM, wavelength-division multiplexing; PC, polarization controller; DP, Dove prism; SPP, spiral phase plate; HW, half-wave plate; BS, 50:50 beam splitter; PBS, polarization beam splitter; MZI, 1-GHz unbalanced Mach-Zehnder interferometers; SPD, single-photon detector with quantum efficiency of $\eta=(10.0\pm 0.2)\%$.}
\label{fig1}
\end{figure*}

\textit{Experiment}-Our experimental setup is descripted in Fig. 1. A narrow-band continuous-wave (cw) laser at 1555.75 nm amplified by an erbium-doped fiber amplifier (EDFA) is frequency doubled in a periodically poled lithium niobate (PPLN) waveguide  by second-harmonic generation (SHG). The remanent pump laser is suppressed by a wavelength division multiplexing (WDM) filter with an extinction ratio of 180 dB. The second harmonic is used to generate photon pairs through the type-0 spontaneous parametric down-conversion (SPDC) process in another 5-cm long PPLN waveguide. Owing to the narrow linewidth of the cw pump, a large emission time uncertainty can be achieved during the SPDC process. Thus one can express the superposition state of the photon pairs emitted at different temporal mode as: $|{\psi}\rangle=\kappa\int_{0}^{\infty}\xi{(t)}|{t}\rangle_{s}|{t}\rangle_{i}dt$, where $ \kappa $ is the coupling constant corresponding to the second-order susceptibility $ \chi^{(2)} $ of the PPLN waveguide, and $ \xi{(t)} $ is the emitted time distribution function. Due to the photon energy conservation during the SPDC, the spectrum of the photon pairs is symmetric with respect to central wavelength of 1555.75 nm and manifests strong frequency correlation, as shown by the SPDC spectrum and joint spectral amplitude (JSA)  in Fig. 2. The SPDC source spans a full width at half maximum (FWHM) of approximate 80 nm, covering the whole telecom C-band and  telecom L-band. We use cascaded 100-GHZ dense wavelength division multiplexing (DWDM) filters to divert the signal and idler photons into 8 standard international telecommunication union (ITU) channels, i.e., ITU CH22 to CH26 for signal and CH28 to CH32 for idler.

\begin{figure}[htbp]
\centering
\includegraphics[width=1\linewidth]{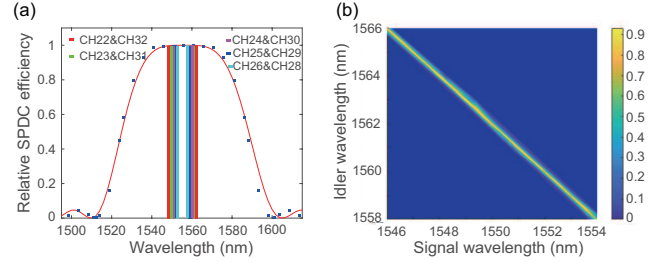}
\caption{(a) Spectrum of the SPDC source based on PPLN waveguide; the red curve is the theoretical result and the blue square points represent experimental data. The colorful bars indicate the time-energy entangled photon pairs multiplexed by 100-GHz DWDM channels. (b) Joint spectral amplitude of signal and idler photons.}
\label{fig2}
\end{figure}

\textit{Results and discussions}-Firstly, we characterize the performance of the SPDC source by using a Franson-type interferometry \cite{Franson1989}, which contains two unequal path-length MZIs with a path delay $ \Delta{t} $ of 1 ns controlled by a piezo-actuated displacement platform. Considering the temporal coherence time of the signal and idler photons to be $ \sigma_{cor}=10 $ ps, such a path delay satisfy the requirement $ \tau\gg\Delta{t}\gg\sigma_{cor} $, where $ \tau $ is the coherence time of the pump laser. Then, the two-photon state combining the OAM freedom is projected to the following form of state:
\begin{equation}
|{\psi}\rangle_{0}=[\frac{1}{\sqrt{2}}(|{t_{1}}\rangle_{s}|{t_{1}}\rangle_{i}+e^{i\varphi}|{t_{2}}\rangle_{s}|{t_{2}}\rangle_{i})]\otimes|{0}\rangle|{0}\rangle,
\label{eq1}
\end{equation}
where $ \varphi $ is the relative phase in the long interferometer arm, and $ |{0}\rangle $ represents that the OAM mode of photon pair is Gaussian mode. In our experiment, another cw laser at the central wavelength of 1570 nm is injected into the other input port of the beam-splitter as feedback to stabilize the phase of interferometers. This reference laser is offset with SPDC photons on optical paths to avoid extra noise. The two-photon interference fringes for the selected photon pairs with respect to the relative phase are shown in Fig. 3. We achieved an average visibility of $ V=(95.1\pm0.5)\% $, revealing a high quality of time-energy entanglement.

\begin{figure}[ht]
\centering
\includegraphics[width=0.7\linewidth]{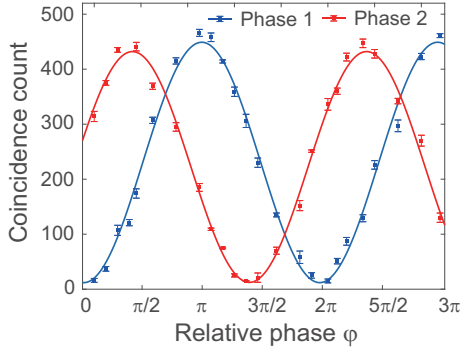}
\caption{Two-photon interference fringes for time-energy entanglement before QET.}
\label{fig3}
\end{figure}

In order to deterministically transfer time-energy entanglement to OAM entanglement, SPPs used as mode-shifters are inserted into the two arms of MZIs to form interferometric quantum gates. When a Gaussian photon transmits through the SPP, its phase acquires a Hilbert factor exp${(i\ell\theta)} $, and its profile becomes OAM mode of $\ell$ ($\ell$=$\pm$1), where $\ell$ is an integer and represents the topological charge. After the OAM mode conversion, the photons passing through the long path carrys an OAM of $\ell_{1}\hbar$ when leaving the interferometer, while the OAM of the other becomes $\ell_{2}\hbar$. Then one can obtain a hyperentangled multi-DOF state:
\begin{equation}
|{\psi}\rangle_{hy}=\frac{1}{\sqrt{2}}(|{t_{1},\ell_{1}}\rangle_{s}|{t_{1},\ell_{1}}\rangle_{i}+e^{i\varphi}|{t_{2},\ell_{2}}\rangle_{s}|{t_{2},\ell_{2}}\rangle_{i}).
\label{eq2}
\end{equation}
After the postselection of photon arrival time and precise adjustment of relative phase $ \varphi $, one can obtain a maximally entangled state $ |{\psi}\rangle=\frac{1}{\sqrt{2}}(|{\ell_{1}}\rangle_{s}|{\ell_{1}}\rangle_{i}+|{\ell_{2}}\rangle_{s}|{\ell_{2}}\rangle_{i})$. To show the flexibility and adaptability of this approach, we experimentally construct four maximally entangled states $ |{\phi^{\pm}}\rangle=\frac{1}{\sqrt{2}}(|{1}\rangle_{s}|{1}\rangle_{i}\pm|{-1}\rangle_{s}|{-1}\rangle_{i})$ and $ |{\psi^{\pm}}\rangle=\frac{1}{\sqrt{2}}(|{1}\rangle_{s}|{-1}\rangle_{i}\pm|{-1}\rangle_{s}|{1}\rangle_{i})$. With an OAM mode of $\ell$$=$1 or $-$1 and a diffraction efficiency $\zeta$ of high than 98\%, four SPPs are inserted within two MZIs to convert the Gaussian photons to OAM-carrying photons. During the experiment, the relative phase of each MZI is carefully control to be $\varphi$=0 or $\pi$. To fully characterize the established OAM states, we perform a quantum state tomography for the four pairs of frequency-correlated channels and reconstruct their density matrices by using the maximum likelihood estimation. Two spatial light modulators (SLM, SLM$_{s}$ for signal photons, SLM$_{i}$ for idler photons) in combination with single mode fibers and single photon detectors are used to characterize the OAM entangled sates, the same as in Ref. \cite{Zhou2016}. The SLMs are used to flatten the spiral phase of incident photons and convert them to a Gaussian mode, which is efficiently coupled to single mode fiber. The real and imaginary parts of the reconstructed density matrices $\rho$ are presented in Fig. 4(a)-(d). We calculate the fidelity relative to the ideal Bell states and obtain the average fidelity $F=\langle{\psi_{ideal}}|\rho|\psi_{ideal}\rangle=(94.1\pm1.3)\% $ and purity $P=Tr(\rho^2)=0.90\pm0.01$, as shown in Fig. 4(f).

\begin{figure}[ht]
\centering
\includegraphics[width=1\linewidth]{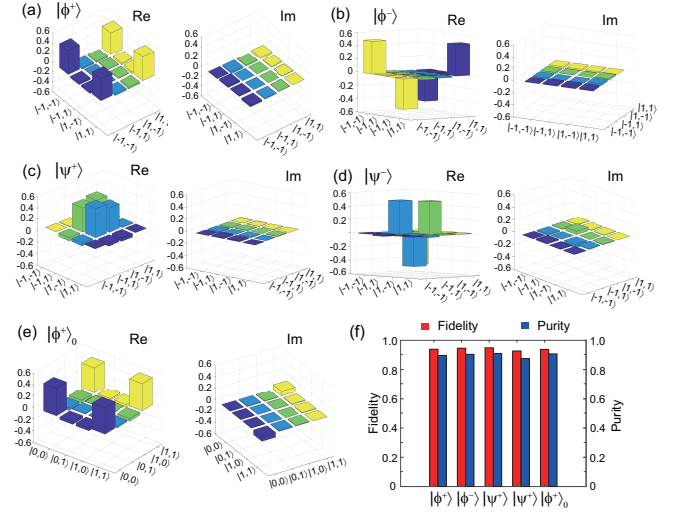}
\caption{(a)-(e) The real and imaginary parts of the reconstructed density matrices correspond to the OAM entangled states $ |{\phi^{\pm}}\rangle=\frac{1}{\sqrt{2}}(|{1}\rangle_{s}|{1}\rangle_{i}\pm|{-1}\rangle_{s}|{-1}\rangle_{i})$, $ |{\psi^{\pm}}\rangle=\frac{1}{\sqrt{2}}(|{1}\rangle_{s}|{-1}\rangle_{i}\pm|{-1}\rangle_{s}|{1}\rangle_{i})$ and  $ |{\phi^{+}}\rangle_{0}=\frac{1}{\sqrt{2}}(|{0}\rangle_{s}|{0}\rangle_{i}+|{1}\rangle_{s}|{1}\rangle_{i})$, respectively. (f) The measured fidelity and purity from the reconstructed density matrices.}
\label{fig4}
\end{figure}

We further measure the two-photon interference fringes to characterize the OAM entangled states. During the measurement, the signal photons are projected onto the state $ |{D}\rangle=\frac{1}{\sqrt{2}}(|{1}\rangle+|{-1}\rangle)$ by SLM$_s$ while the idler photons onto $ |{\theta}\rangle=\frac{1}{\sqrt{2}}(e^{i\theta}|{1}\rangle+e^{-i\theta}|{-1}\rangle)$ by SLM$_i$. Then, two-photon coincidences are recorded as a function of the rotation angle $\theta$ of the phase mask applied to SLM$_i$. By scanning $\theta$ from 0 to 2$\pi$, the two-photon interference patterns are obtained, as shown in Fig. 5. The average fringe visibility for four OAM entangled states is calculated to be \textit{V}=(95.4$\pm$1.8)\%, which exceeds the 71\% local bound of the Bell's inequality and convincingly reveals the existence of entanglement. The experimental results powerfully confirm the quality of the two-photon entanglement in OAM DOF after QET by using such an interferometric gate.

\begin{figure}[ht]
\centering
\includegraphics[width=0.8\linewidth]{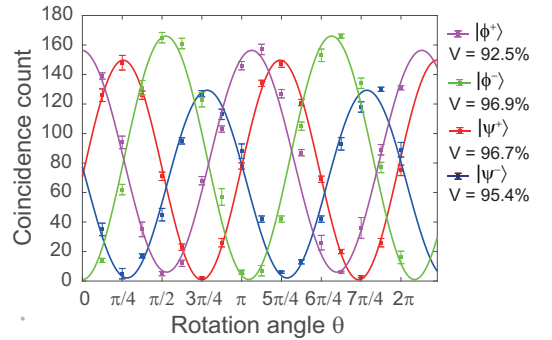}
\caption{Two-photon interference fringes of four OAM entangled states for observing the violation of Bell inequality. Each coincidence count is measured for 10 seconds.}
\label{fig5}
\end{figure}

To prove the ability of the two-way QET, we next demonstrate that two-photon OAM entangled state can be deterministically transferred to time-energy entangled sate by using a interferometric cyclic gate. Compared with QET from time-energy to OAM DOF, it is more challenging to coherently convert entanglement in OAM DOF back to time-energy subspace. The polarization of signal and idler photons is rotated into $\frac{1}{\sqrt{2}}( |H\rangle+|V\rangle)$ by two half-wave plates (HW) before they enter the interferometric cyclic gates. Then, a double-path Sagnac interferometer containing two DPs is utilized to serve as an OAM sorter to split different OAM modes. Considering that the quantum state of the incident photons is the superposition of two arbitrary OAM modes, $\ell_1$ and $\ell_2$, the polarizations of photons with different modes will pose perpendicularly to each other \cite{Zhang2014} once the relative orientation of two DPs is $\alpha$=\ang{90}$/(\ell_{1}-\ell_{2})$. In our experiment, we choose the quantum state $ |{\phi^{+}}\rangle_{0}=\frac{1}{\sqrt{2}}(|{0}\rangle_{s}|{0}\rangle_{i}+|{1}\rangle_{s}|{1}\rangle_{i})$ as an example to demonstrate the QET from OAM DOF to time-energy subspace. We firstly prepare the state $ |{\phi^{+}}\rangle_{0}$ through QET from time-energy to OAM DOF. A SPP with OAM mode of $\ell$=1 is inserted in the short path of each MZI of the first Franson-type interferometer. At this time, the SPPs in the long path of MZIs shown in the OAM preparation part of Fig. 1 are removed. After reconstructing the density matrix, we obtain the fidelity of 93.8\% and state purity of 90.7\%, as shown in Fig. 4(e) and (f). To further characterize the entanglement of the prepared state, we measure the \textit{S} parameter of the Clauser-Horne-Shimony-Holt (CHSH) inequality as in Ref. 26 and obtain the result of \textit{S}=2.4$\pm$0.03, which violates the inequality by 13 standard deviations.

\begin{figure}[ht]
\centering
\includegraphics[width=0.7\linewidth]{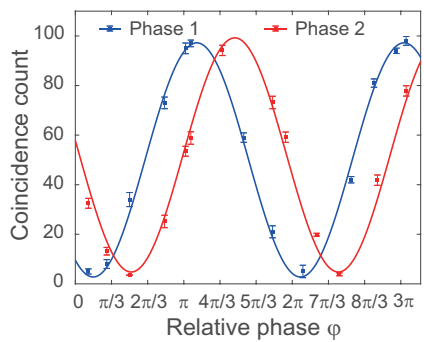}
\caption{Two-photon interference fringes for time-energy entanglement after QET.}
\label{fig6}
\end{figure}

Then, we set the relative orientation of the two DPs in the Sagnac interferometer to be $\alpha=90^{0}$. At the output of the Sagnac interferometer, the photons of different modes are separated into different paths by polarization beam splitter 1 (PBS1). Photons with an OAM mode of +1 take the long path, when the others pass through the short path. The two paths are recombined by PBS2, forming a MZI with a delay time of 1 ns. In order to erase any information about OAM profile, a SPP and a 4-f system are inserted in the MZI to convert the OAM photons to the Gaussian mode. Finally, the two-photon state becomes a time-energy entangled state after the erasing of information about polarization and path, then it can be collected into fiber for long distance distribution. We use another Franson-type interferometry consisting of two MZIs with 1-ns relative delay to reveal the generated time-energy entangled state. The experimental results are shown in Fig. 6. We obtain a fitted visibility of \textit{V}=(92.4$\pm$1.6)\% for the time-energy entangled state, which implies that the quantum entanglement is coherently transferred into time-energy DOF.

The most important feature of our quantum interface is that it maintains the quantum characteristics of the output photons after QET. In our experiment, the entangled state  can maintain a high fidelity over 90\% for both QET processes. However, the degradation of the visibility of interference fringes for time-energy entanglement is still non-negligible after QET. This is mainly caused by the imperfection of optical elements, system loss due to non-unity coupling efficiency of space to fiber, and low detection efficiency of SPDs, which can be improved by optimizing system parameters and using a high-performance SPD \cite{Marsili2013}. Another key feature is the high probability of success of entanglement transfer. With the help of postselection of photon arrival time, a deterministic entanglement transfer from time energy to OAM DOF can be theoretically achieved. In our experiment, the probability of success of entanglement conversion is about 96\%. As for the other process, due to the high extinction ratio of OAM sorters and the single-mode characteristics of fiber, the QET of OAM back to time-energy DOF is deterministic, too. On the other hand, our proposed QET interface has the following preponderant characteristics. Firstly, our method paves the way for preparing a multi-channel OAM entangled source by using DWDM technology, which is difficult to implement by directly pumping a nonlinear crystal. In addition, an arbitrary two dimensional OAM entangled state can be prepared by replacing the first beam splitter in each MZI of the Franson-type interferometer with a PBS and a HW. Secondly, the time-energy and OAM DOF belong to high-dimensional Hilbert space. The interferometric cyclic gates can be cascaded with each other to achieve high-dimensional QET, which is promising in high-dimensional quantum tasks.

\textit{Summary}-In conclusion, we have demonstrated a two-way QET interface with high fidelity between the time-energy and the OAM DOF. Based on this interface, we firstly implement the QET from time-energy to OAM DOF with an average fidelity of the OAM entangled states of higher than 94.1\%. Then, we show quantum entanglement can be coherently tranferred from OAM back to time-energy DOF with a high visibility of Franson-type interference over 92.4\%. This interface can be used to prepare multi-channel OAM entangled sources and paves a new way for establishing entanglement between remote heterogeneous quantum nodes. Thus, our scheme has great potential applications in future quantum communication network, such as multi-DOF quantum entanglement swapping and quantum direct communication on multinode integrated space-to-fiber communication networks.

 \begin{acknowledgments}
This work is supported in part by the National Key Research and Development Program of China (Grant No. 2017YFA0303700), National Natural Science Foundation of China (Grant Nos. 11734011, 11804135, and 12074155), The Foundation for Shanghai Municipal Science and Technology Major Project (Grant No. 2019SHZDZX01-ZX06), and Project funded by China Postdoctoral Science Foundation (Grant No. 2019M661476).
\end{acknowledgments}

\nocite{*}

\bibliography{reference}

\end{document}